\documentclass[prl,twocolumn,superscriptaddress,preprintnumbers,amsmath,amssymb]{revtex4-2}
\usepackage{graphicx}
\usepackage{dcolumn}
\usepackage{amsmath,amsbsy,amssymb,scalefnt}
\usepackage{bm}
\usepackage{epstopdf}
\usepackage[T1]{fontenc}
\usepackage{amsfonts}
\usepackage{booktabs}
\usepackage{colortbl}
\usepackage{textcomp}
\usepackage{mathtools}
\usepackage{xcolor}
\usepackage{hyperref}
\usepackage{physics}
\hypersetup{ pdfnewwindow=true, colorlinks=true, linkcolor=blue, anchorcolor=blue, citecolor=blue, filecolor=blue, menucolor=blue, urlcolor=blue}

\begin{document}
\title{Linearly polarized light enables chiral edge transport in quasi-2D Dirac materials}

\author{Mohammad Shafiei}
\affiliation{COMMIT, Department of Physics \& NANOlight Center of Excellence, University of Antwerp, Groenenborgerlaan 171, B-2020 Antwerp, Belgium}
\author{Farhad Fazileh}
\affiliation{Department of Physics, Isfahan University of Technology, Isfahan 84156-83111, Iran}
\author{Milorad V. Milo\v{s}evi\'c}
\email{milorad.milosevic@uantwerpen.be}
\affiliation{COMMIT, Department of Physics \& NANOlight Center of Excellence, University of Antwerp, Groenenborgerlaan 171, B-2020 Antwerp, Belgium}

\date{\today}

\begin{abstract}
Floquet engineering with high-frequency light offers dynamic control over topological phases in quantum materials. While in 3D Dirac systems circularly polarized light is known to induce topological phase transitions via gap opening, linearly polarized light (LPL) has generally been considered ineffective. Here we show that in quasi-2D Dirac materials the second-order momentum term arising from the intersurface coupling can induce a topological phase transition under LPL, leading to chiral edge channels. Considering an ultrathin Bi$_2$Se$_3$ film as a representative system, we show that this transition occurs at experimentally accessible light intensities. Our results thus promote quasi-2D materials as viable platforms for light-controlled topological phases, expanding the potential of Floquet topological engineering.
\end{abstract}

\maketitle

\paragraph{Introduction}
Quasi-two-dimensional (quasi-2D) materials are known as versatile platforms to explore light–matter interactions and engineer tunable quantum phases~\cite{bao2022light, tao2021enhancing, huang2017emerging}. Their reduced dimensionality, enhanced surface-to-volume ratio, and strong susceptibility to external perturbations~\cite{shafiei2024tuning} make them ideal for realizing novel electronic and topological states~\cite{bian2022strong}. In particular, intersurface coupling intrinsic to quasi-2D systems induces tunable bandgaps and hybridization effects absent in their bulk counterparts~\cite{zhang2010crossover}, offering promising opportunities for next-generation optoelectronic and spintronic technologies~\cite{zhao2017high, das2021transistors}.

Floquet engineering is a particularly powerful strategy to dynamically tailor the electronic structure of such quasi-2D materials, where time-periodic driving, typically via high-frequency light, modifies the band topology of the system out of equilibrium~\cite{bao2022light,koo2024dynamical,liu2023floquet,xu2021light}. Within this framework, photon energies exceeding characteristic electronic scales allow for band inversion, Floquet gap formation, and topological phase transitions~\cite{rudner2020band, dabiri2021engineering, dehghani2016occupation}. Such light-induced control has led to the realization of a variety of exotic states, including Floquet topological insulators (TIs) and quantum anomalous Hall phases~\cite{zhang2025floquet, shafiei2024floquet}.

Most of these developments have relied on circularly polarized light (CPL), which inherently breaks time-reversal symmetry and opens topological gaps at Dirac points~\cite{dabiri2021light}. In contrast, linearly polarized light (LPL) has generally been considered ineffective for inducing topological edge states or bandgap openings in three-dimensional or bulk systems~\cite{choudhari2019effect}. While LPL could reshape band dispersions~\cite{shafiei2025light,choudhari2019effect}, it has not been expected to drive topological phase transitions.

In this work, we challenge this prevailing view by demonstrating that LPL can indeed induce topological phases in quasi-2D Dirac materials. We show that second-order momentum contributions, arising from the strong intersurface coupling in quasi-2D systems, enable a unique Floquet mechanism that is absent in 3D analogues. The light field interacts non-trivially with intersurface coupling, generating a dynamical massive Dirac cone. Using ultrathin Bi$_2$Se$_3$ films as a prototypical system, we reveal the emergence of chiral edge channels under high-frequency LPL, previously thought to require CPL for experimentally accessible light intensities. Our results thus redefine the role of LPL in Floquet engineering and significantly broaden the range of experimentally accessible topological phases in driven quantum materials.

\paragraph{System and Hamiltonian}

We consider a quasi-two-dimensional Dirac system modeled by an effective low-energy Hamiltonian that incorporates both linear and quadratic momentum dependencies, arising from intersurface coupling and confinement effects. The Hamiltonian is given by:
\begin{equation}
H(\mathbf{k}) = v_F (k_y \sigma_x - \alpha\, k_x \sigma_y) + \left( \Delta_0 - \Delta_1 k^2 \right) \sigma_z,
\end{equation}
where \( \sigma_i \) are the Pauli matrices acting in pseudospin space, \( k^2 = k_x^2 + k_y^2 \), \( v_F \) is the Dirac velocity, and \( \alpha \) accounts for possible anisotropy. The term \( \left( \Delta_0 - \Delta_1 k^2 \right) \sigma_z \) introduces a momentum-dependent mass gap, where \( \Delta_0 \) and \( \Delta_1 \) are material-specific parameters that can be extracted from experimental data. 

While our theoretical framework applies to any quasi-2D Dirac system exhibiting quadratic corrections to the mass term, we focus specifically on ultrathin Bi$_2$Se$_3$ films, for which strong intersurface hybridization leads to finite \( \Delta_0 \), \( \Delta_1 \), and \( \alpha = 1 \), thus providing a realistic platform for studying Floquet-induced topological phase transitions under experimentally accessible conditions.

Bi$_2$Se$_3$ is a prototypical three-dimensional TI with a rhombohedral crystal structure belonging to the $R\bar{3}m$ space group~\cite{zhang2009topological}. Each unit cell comprises a quintuple layer (QL) consisting of two bismuth (Bi) and three selenium (Se) atoms, stacked along the $z$-axis~\cite{zhang2009topological,zhang2010crossover}. In ultrathin films comprising a few QLs, hybridization between the top and bottom surface states opens a finite energy gap at the Dirac point, known as the hybridization gap, captured effectively by the parameters \( \Delta_0 \) and \( \Delta_1 \) in the low-energy Hamiltonian. These parameters are thickness-dependent and rapidly decrease with increasing film thickness, eventually vanishing beyond approximately five QLs, where surface states decouple and the system approaches the gapless limit of a bulk TI. The experimentally extracted values of \( \Delta_0 \), \( \Delta_1 \), and \( v_F \) are summarized in Table~\ref{tab:params}~\cite{zhang2010crossover}.

\begin{table}[b]
\centering
\begin{tabular}{|c|c|c|c|c|c|}
\bottomrule
\bottomrule
& \textbf{2QL} & \textbf{3QL} & \textbf{4QL} & \textbf{5QL} & $\geq$ \textbf{6QL} \\
\midrule
$\Delta_0$ (eV) & 0.126 & 0.069 & 0.035 & 0.020 & 0 \\
$\Delta_1$ (eV$\cdot$\AA$^2$) & 21.8 & 18 & 10 & 5 & 0 \\
$v_F$ (eV$\cdot$\AA) & 3.10 & 3.17 & 2.95 & 2.99 & 2.98 \\
\bottomrule
\bottomrule
\end{tabular}
\caption{Thickness dependence of the parameters $\Delta_0$, $\Delta_1$, and $v_F$ for ultrathin Bi$_2$Se$_3$ films (2--5 QLs) and thick films ($\geq$6 QLs). Data are extracted from experimental measurements~\cite{zhang2010crossover}.}
\label{tab:params}
\end{table}

\begin{figure}[b]
    \centering
    \includegraphics[width=0.98\linewidth]{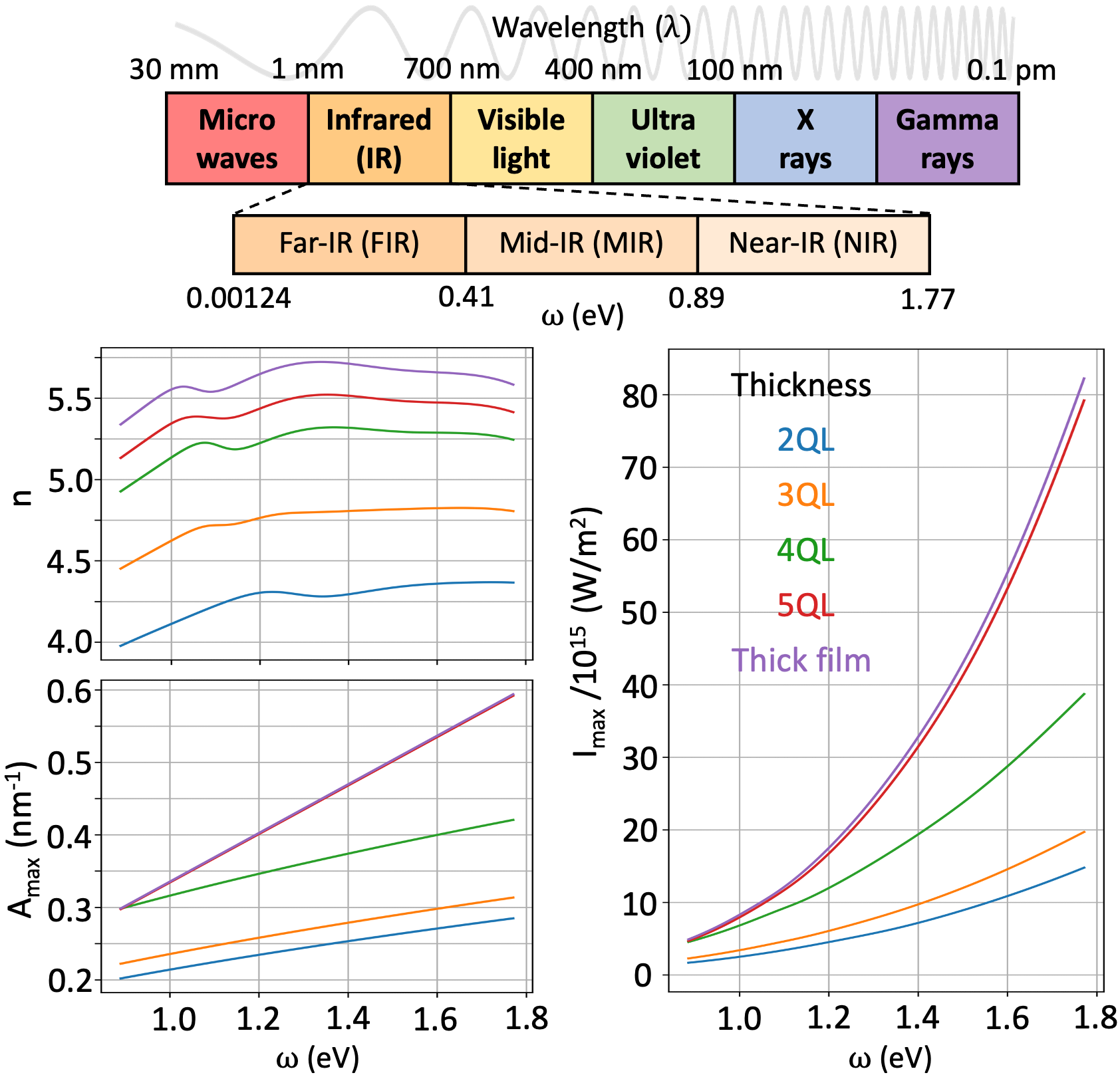}
    \caption{Electromagnetic spectrum is shown with corresponding wavelength and angular frequency ranges. NIR region is selected to satisfy the high-frequency condition relative to the $\sim$0.3~eV bandgap of Bi$_2$Se$_3$, enabling effective Floquet engineering. Plotted are the refractive index (extracted from Ref.~\cite{fang2020layer}), and the maximum permissible optical field amplitude and intensity, for Bi$_2$Se$_3$ films of various thicknesses (2–5 QLs as ultrathin film and 8 QLs as a representative thick film) within the NIR regime.}
    \label{fig:fig1}
\end{figure}

To investigate the influence of periodic driving, we focus on the high-frequency regime where the photon energy \( \hbar\omega \) exceeds all other relevant energy scales, including the bandgap. Given that the bulk bandgap of Bi$_2$Se$_3$ is approximately 0.3~eV~\cite{zhang2009topological}, we consider near-infrared (NIR) light sources with photon energies ranging from 0.89~eV to 1.77~eV (assuming \(\hbar = e = 1\)). The corresponding electromagnetic spectrum is depicted in Fig.~\ref{fig:fig1}.

We model the incident optical field as a linearly polarized electric field propagating along the $z$-axis, perpendicular to the plane of the film. The time-dependent electric field is derived from a vector potential \( \mathbf{A}(t) \) as:
\begin{equation}
\mathbf{E}(t) = -\frac{\partial \mathbf{A}(t)}{\partial t} = E_0 \left( \cos(\omega t), \cos(\omega t) \right),
\end{equation}
where \( E_0 \) is the electric field amplitude and \( \omega \) is the angular frequency of the incident light.

The interaction of the light field with the electronic system is incorporated through the Peierls substitution, \( \mathbf{k} \to \mathbf{k} + \mathbf{A}(t) \). We neglect the Zeeman coupling between the Dirac spin and the light-induced magnetic field, as even for electric field strengths as high as \( E_0 \sim 10^9 \) V/m, the associated magnetic field (\( B_0 \sim 10 \) T) generates a negligible Zeeman gap of \(\sim 0.001\)~eV~\cite{choudhari2019effect}, much smaller than the intrinsic energy scales of the system.

The time evolution of the system is governed by the time-dependent Schrödinger equation:
\begin{equation}\label{schro}
H(t) \Psi(\mathbf{r}, t) = i\hbar \frac{\partial}{\partial t} \Psi(\mathbf{r}, t),
\end{equation}
where \( \mathbf{r} = (x, y) \). Due to the periodicity of \( H(t) \) with period \( T = 2\pi/\omega \), we employ the Floquet formalism, expanding the wavefunction as~\cite{kitagawa2011transport,choudhari2019effect}:
\begin{equation}
\Psi(\mathbf{r}, t) = \sum_{m=-\infty}^{\infty} e^{-i(\epsilon + m\hbar\omega)t/\hbar} u^m(\mathbf{r}),
\end{equation}
where \( \epsilon \) is the quasienergy and \( u^m(\mathbf{r}) \) are the time-independent Floquet modes. Substituting into Eq.~\eqref{schro} and projecting onto the Floquet basis yields an effective time-independent Hamiltonian as:
\begin{equation}
H_{\text{eff}}(\mathbf{k}) = \left( \Delta_0 - \Delta_1 k^2 - \frac{\Delta_1 A^2}{2} \right) \sigma_z + v_F (k_y \sigma_x - k_x \sigma_y),
\end{equation}
where \( A = E_0/\omega \) is the amplitude of the vector potential.

\begin{figure*}
    \centering
    \includegraphics[width=0.85\linewidth]{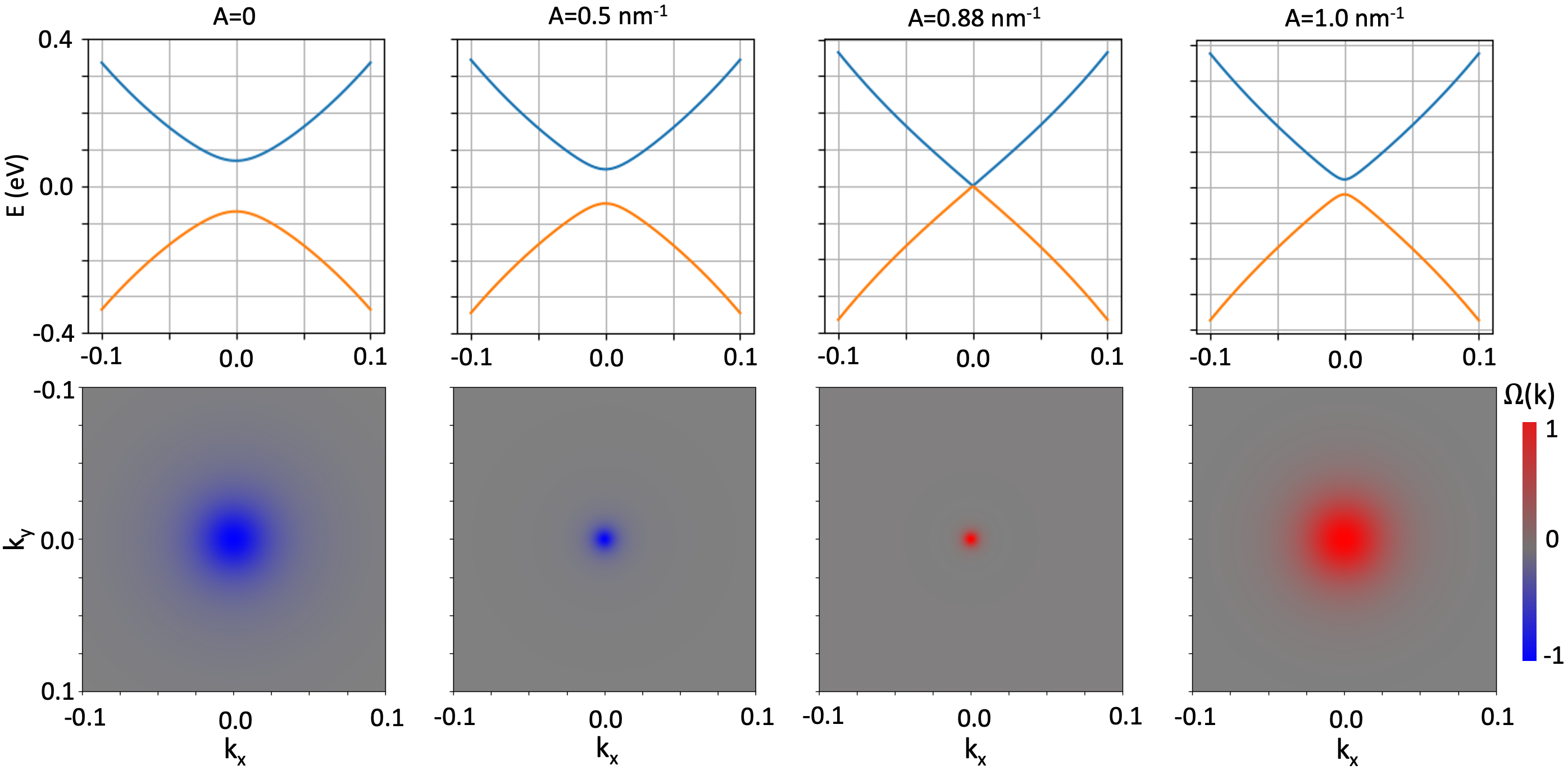}
    \caption{Floquet band structure and Berry curvature of a 3QL Bi$_2$Se$_3$ film under LPL. The applied light induces a bandgap modulation in the surface states, leading to a topological phase transition. The sign reversal of the Berry curvature at critical field amplitude ($A = 0.88$~nm$^{-1}$) confirms the emergence of a light-induced topological phase.}
    \label{fig:fig2}
\end{figure*}

\paragraph{Results and Discussion}

To validate the applicability of the high-frequency approximation used in our Floquet analysis, we first evaluate the energy scale associated with the time-dependent Hamiltonian \( H(\mathbf{k}, t) \) averaged over one period \( T = 2\pi/\omega \) of the driving field. Focusing on the \(\Gamma\) point (\(\mathbf{k} = 0\)), where the instantaneous energy is maximized, we impose the condition~\cite{qin2024light}:
\begin{equation}
\frac{1}{T} \int_0^T dt\, \| H(0, t) \|_{\text{max}} \lesssim \omega,
\end{equation}
ensuring that the system remains within the regime where the high-frequency expansion is reliable. 
Including the light–matter interaction through minimal coupling (\(\mathbf{k} \to \mathbf{k} + e\mathbf{A}(t)\)) and keeping dominant terms, we derive an upper limit for the amplitude \(A\) of the vector potential: \(A_{\text{max}} = \min\left(\omega/v_F, \sqrt{\omega/\Delta_1}\right)\)~\cite{qin2024light}. Here, $A$ characterizes the optical field strength, and the corresponding intensity is given by \(I = \frac{1}{2} n c \varepsilon_0 |A \omega/e|^2\), with \(n\) the refractive index, \(c\) the speed of light, and \(\varepsilon_0\) the vacuum permittivity~\cite{paschotta2008encyclopedia}.

Figure~\ref{fig:fig1} shows the refractive index spectra for Bi\(_2\)Se\(_3\) thin films of different thicknesses (2–5 QLs and 8 QLs) in NIR regime, adapted from Ref.~\cite{fang2020layer}, alongside the calculated maximum optical field amplitudes and intensities as a function of frequency. We will show that the light amplitudes and therefore intensities required to achieve Floquet engineering of the band structure lie well below the thresholds imposed by the high-frequency condition, thereby confirming the feasibility of experimental realization of our findings.

To examine the phase transition, we first analyze the Floquet band structure of an ultrathin TI film subjected to LPL. The quasienergy spectrum and Berry curvature distribution for a 3QL Bi\(_2\)Se\(_3\) film are shown in Fig.~\ref{fig:fig2}. As the optical field amplitude increases, a band inversion occurs at the Dirac point, accompanied by a change in the sign of the Berry curvature. The Berry curvature \(\Omega(\mathbf{k})\) is computed from the Floquet eigenstates \(|u(\mathbf{k})\rangle\) through:
\begin{equation}
\Omega(\mathbf{k}) = i \left( \left\langle \frac{\partial u}{\partial k_x} \bigg| \frac{\partial u}{\partial k_y} \right\rangle - \left\langle \frac{\partial u}{\partial k_y} \bigg| \frac{\partial u}{\partial k_x} \right\rangle \right).
\end{equation}
The sign reversal of \(\Omega(\mathbf{k})\) at a critical amplitude \(A = 0.88\,\mathrm{nm}^{-1}\) signals a light-induced topological phase transition.

\begin{figure}[b]
    \centering
    \includegraphics[width=0.9\linewidth]{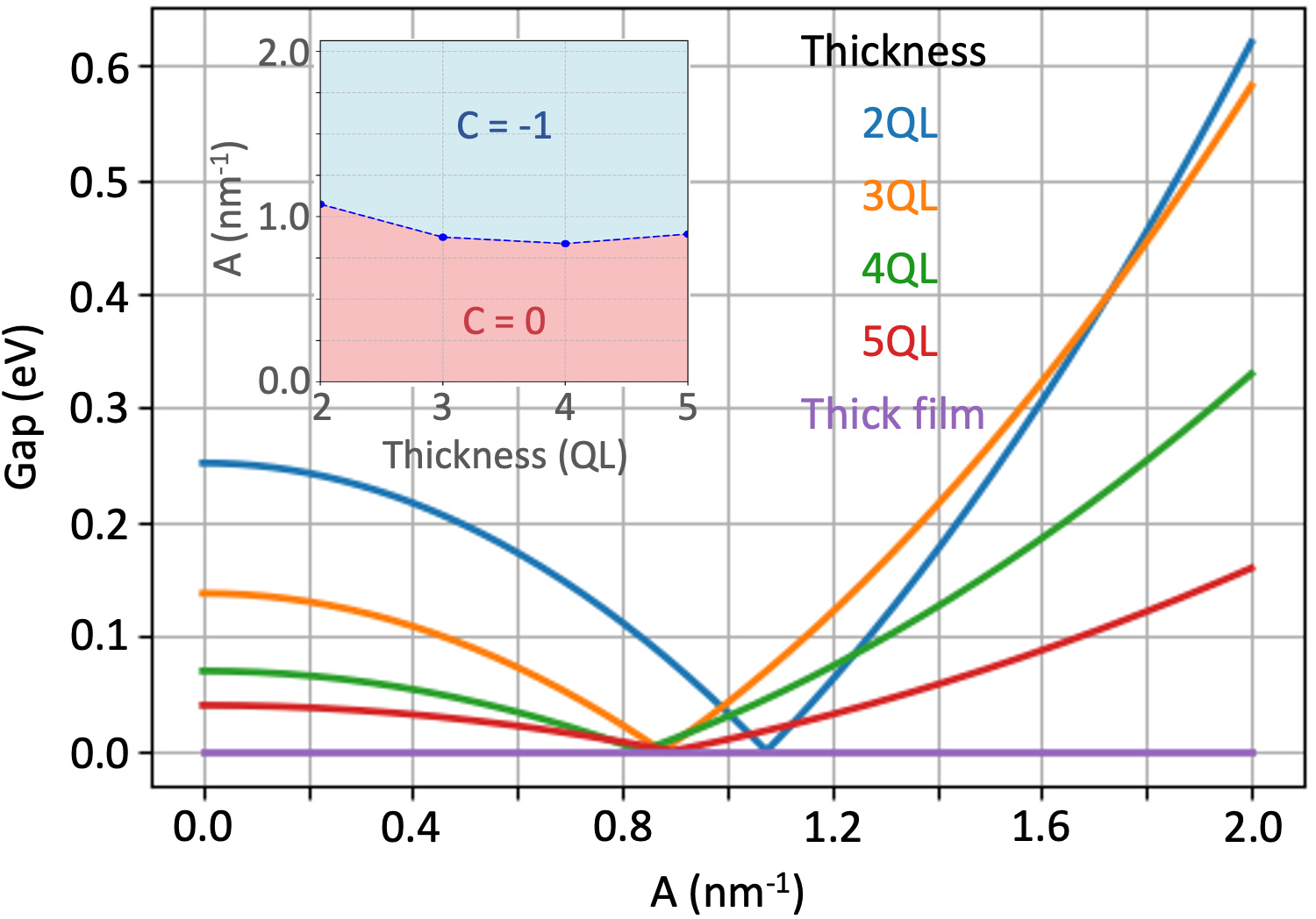}
    \caption{Light-induced modulation of the surface-state bandgap in Bi$_2$Se$_3$ ultrathin (2-5 QLs) and thick films under LPL. Bandgap variation versus optical field amplitude $A$ exhibits gap closing and reopening, indicating a topological phase transition. The inset shows the threshold amplitude $A_{\text{th}}$ for each thickness in the ultrathin regime. The thick film remains topologically unaffected by the applied field, in the entire range considered.}
    \label{fig:fig3}
\end{figure}

To systematically map the topological response, we studied the evolution of the bandgap in Bi\(_2\)Se\(_3\) films of thickness 2–5 QLs, as well as in a thick film, as a function of the optical field amplitude, as shown in Fig.~\ref{fig:fig3}. For ultrathin films, the gap closes and reopens as the amplitude crosses a critical threshold, indicative of a topological transition. In contrast, a thick film, where hybridization between top and bottom surfaces is negligible, exhibits no gap opening/closing under similar illumination conditions.

Quantitative analysis reveals that the threshold field amplitude \(A_{\text{th}}\) required for the transition depends on the intrinsic hybridization parameter \(\Delta_1\) that depends on the thickness of the film. Threshold values are found to be approximately 1.07, 0.88, 0.83, and 0.89\,nm\(^{-1}\) for 2-5QL films, respectively. The phase transition corresponds to a change in the Chern number from \(C = 0\) (normal insulator) to \(C = -1\) (Chern insulator), enabling the emergence of dissipationless chiral edge channels. The Chern number $C$ for a particular band is calculated as an integral over the Brillouin zone~\cite{tokura2019magnetic}:
\begin{equation}
    C = \frac{1}{2\pi}\int_{BZ} \Omega (k)\, dk.
\end{equation}

\begin{figure}[b]
    \centering
    \includegraphics[width=0.85\linewidth]{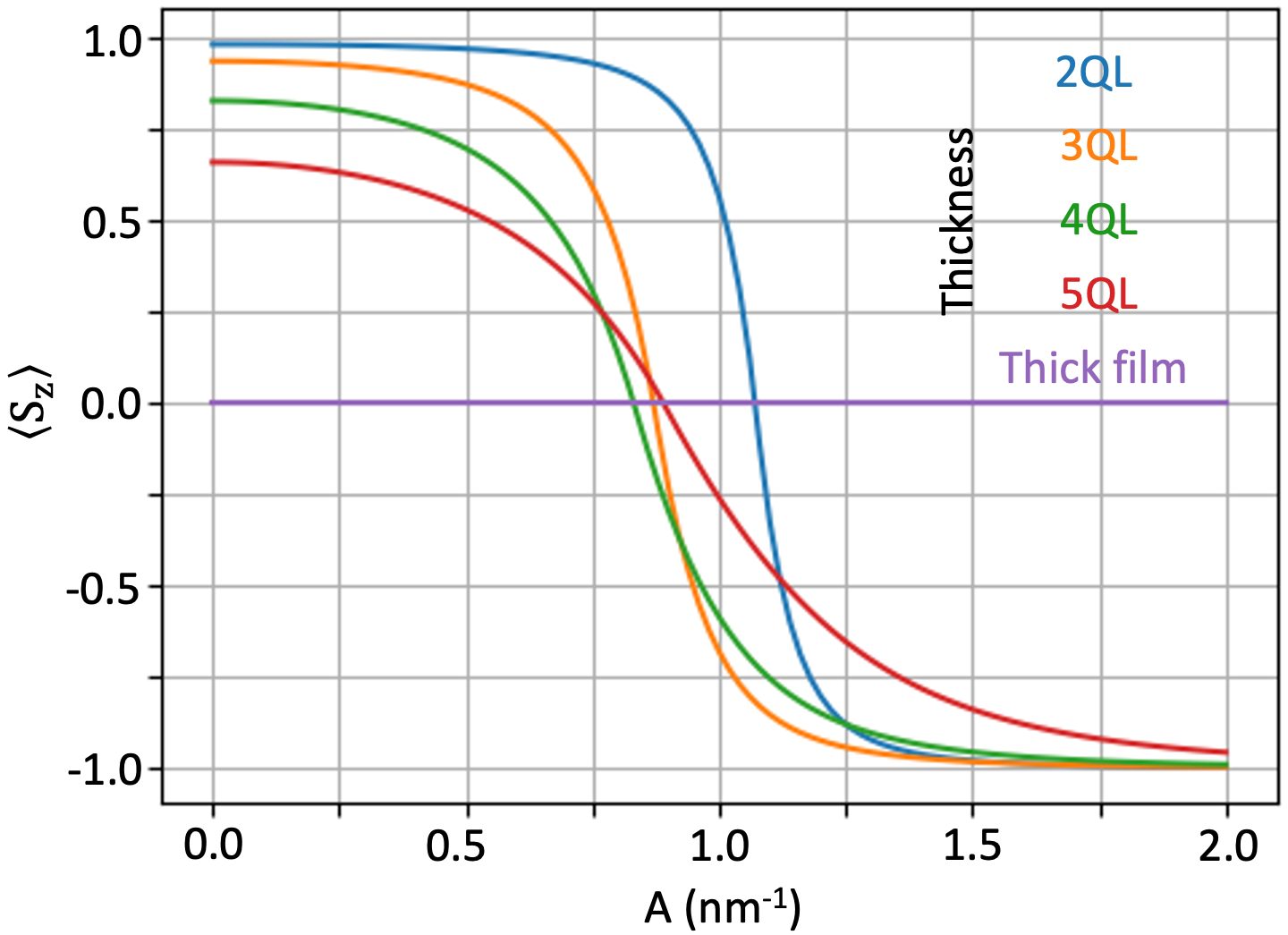}
    \caption{Average $z$-component of the spin expectation value $\langle S_z \rangle$ versus optical field amplitude $A$, for ultrathin and thick Bi$_2$Se$_3$ films. While $\langle S_z \rangle$ remains zero for the thick film due to the preserved helical spin texture, it varies non-monotonically in ultrathin films, changing sign at a threshold amplitude $A_{\text{th}}$, indicating a light-induced topological phase transition.}
    \label{fig:fig4}
\end{figure}

Further insight into the transition is provided by monitoring the behavior of the spin texture. The spin texture is the vector field in momentum space representing the expectation values of the spin orientations, calculated as:
\begin{equation}
\vec{S} = \frac{1}{2} \langle u_k | \vec{\sigma} | u_k \rangle = \frac{1}{2}[\langle\sigma_x\rangle \hat{x} + \langle\sigma_y\rangle \hat{y} + \langle\sigma_z\rangle \hat{z}].
\end{equation}
The topological phase transition is then corroborated by analyzing the average $z$-component of the spin expectation value, $\langle S_z \rangle$, as a function of the optical field amplitude, as presented in Fig.~\ref{fig:fig4}. For thick films, \(\langle S_z \rangle\) remains zero at all light intensities, consistent with preserved helical spin textures and time-reversal symmetry. In contrast, ultrathin films exhibit a nonmonotonic evolution of $\langle S_z \rangle$, including a sign reversal at $A = A_{\text{th}}$, consistent with the onset of a Floquet-induced topological phase.

\begin{figure}[b]
    \centering
    \includegraphics[width=0.9\linewidth]{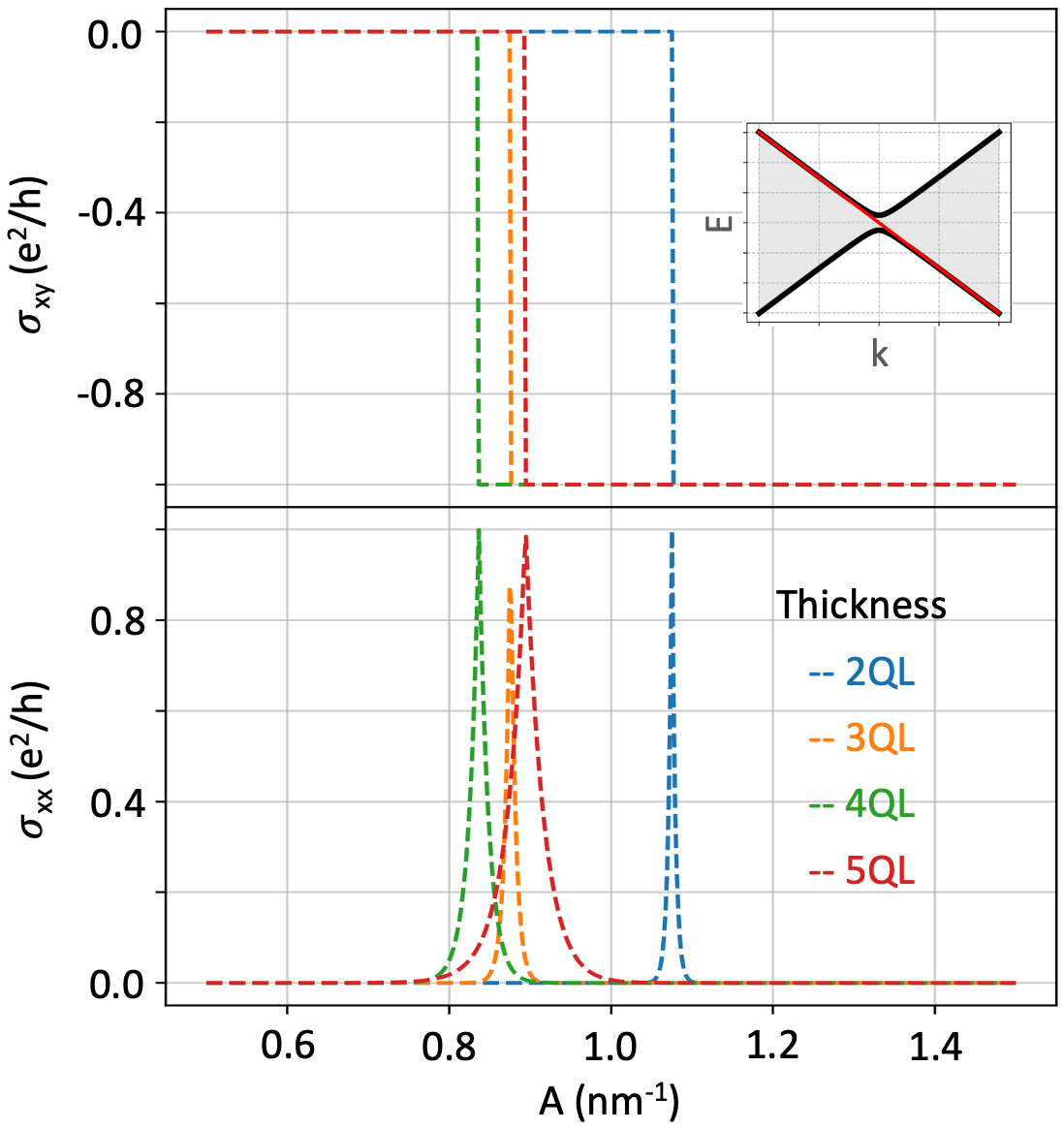}
    \caption{Longitudinal ($\sigma_{xx}$) and Hall ($\sigma_{xy}$) conductances of Bi$_2$Se$_3$ ultrathin films (2–5 QLs) under LPL as a function of the optical field amplitude $A$, with $E_F = 0.001$~eV and temperature $\Theta = 0.001$. The quantized Hall plateau $\sigma_{xy} = -e^2/h$ and vanishing $\sigma_{xx}$ for $A > A_{\text{th}}$ confirm the emergence of chiral edge states with Chern number $C = -1$.}
    \label{fig:fig5}
\end{figure}

Finally, we compute the longitudinal and Hall conductivities, \(\sigma_{xx}\) and \(\sigma_{xy}\), as functions of the optical field amplitude. 
We compute the longitudinal and Hall conductivities using the Kubo framework, expressed as~\cite{sinitsyn2006charge,streda1982theory}:
\begin{equation}
    \sigma_{ij} = - \frac{e^2}{2h} \int d\omega \frac{\partial f(\omega)}{\partial \omega} \sum_{\mathbf{k}} \text{Tr} \big\{ v_i [ G^r (\mathbf{k}, \omega) - G^a (\mathbf{k}, \omega) ] \big.
\end{equation}
\begin{equation*}
    \big. \times v_j G^a (\mathbf{k}, \omega) - v_i G^r (\mathbf{k}, \omega) v_j [ G^r (\mathbf{k}, \omega) - G^a (\mathbf{k}, \omega) ] \big\},
\end{equation*}
where \( f(\omega) \) is the Fermi-Dirac distribution, \( v_i \) is the velocity operator, and \( G^{r/a}(\mathbf{k}, \omega) \) are the retarded/advanced Green’s functions. These are decomposed as:
\begin{equation}
G^{r/a}(\mathbf{k}, \omega) = \frac{1}{2} \sum_{\gamma = \pm} \left[ 1 + \gamma\, \mathbf{n_k} \cdot \bm{\sigma} \right] G^{r/a}_\gamma(\mathbf{k}, \omega),
\end{equation}
with \( \mathbf{n_k} = (-k_y, k_x, 0)/|\mathbf{k}| \) denoting the momentum unit vector.

As shown in Fig.~\ref{fig:fig5}, a quantized Hall plateau \(\sigma_{xy} = -e^2/h\) emerges for \(A > A_{\text{th}}\), while \(\sigma_{xx}\) simultaneously vanishes, signifying the formation of a robust chiral edge mode. Notably, \(\sigma_{xx}\) exhibits a sharp peak at \(A = A_{\text{th}}\), corresponding to the transient gapless point at the topological transition. Due to the structure of the effective Hamiltonian, where the induced mass term scales as $-\Delta_1 A^2/2$, only the $C = -1$ phase is accessible in this regime. These quantized transport responses serve as unambiguous indicators of the transition and offer a viable route for experimental validation through optical pumping and magnetotransport measurements.

\paragraph{Conclusion}
In summary, we have demonstrated that linearly polarized high-frequency light, previously considered ineffective for inducing topological transitions in Dirac systems, can drive a Floquet-engineered topological phase transition in quasi-two-dimensional Dirac materials. This transition stems from the second-order momentum-dependent terms due to intersurface hybridization, which become significant in ultrathin samples. Focusing on an ultrathin Bi$_2$Se$_3$ film, we predict a Floquet-induced Chern insulating phase characterized by quantized Hall conductance and robust chiral edge states, achieved without magnetic doping or circularly polarized light. In addition, the critical optical field strength required for the topological transition lies within the experimentally accessible limits, underscoring the practical feasibility of realizing such phases.

Our findings reveal a previously unexplored route to light-induced topological states via linear polarization, broadening the applicability of Floquet engineering and its integration with optoelectronic platforms. The results also highlight the crucial impact of dimensional confinement and interlayer coupling on the non-equilibrium band structure.
By positioning quasi-2D topological insulators as versatile hosts for optically tunable topological phases, this work opens additional pathways toward ultrafast, dissipationless devices for quantum information processing, spintronics, and reconfigurable photonic systems.

\paragraph{Acknowledgments} This research was supported by the Research Foundation-Flanders (FWO-Vlaanderen), the FWO-FNRS EOS project ShapeME, and the Special Research Funds (BOF) of the University of Antwerp.

\bibliography{bibliography}

\end{document}